\documentclass[epj]{svjour}
\usepackage{graphics}
\usepackage{dcolumn}
\usepackage{amsmath}
\usepackage{amssymb}
\usepackage{graphics}
\usepackage{epstopdf}
\usepackage{bm}% bold math
\newcommand{\bea}{\begin{eqnarray}}
\newcommand{\eea}{\end{eqnarray}}
\newcommand{\be}{\begin{equation}}
\newcommand{\ee}{\end{equation}}
\newcommand{\f}[1]{\frac{1}{2}}

\def\be{\begin{eqnarray}}
\def\ee{\end{eqnarray}}
\def\bd{\begin{displaymath}}
\def\ed{\end{displaymath}}

\def\sl{^-\!\!\!\!}
\def\etal{{\em et al.}}
\def\NP{Nucl. Phys. }
\def\PR{Phys. Rev. }
\def\PRL{Phys. Rev. Lett. }

\def\jpg{J. Phys. G: Nucl. Part. Phys. }
\def\epj{Eur. Phys. J. }

\begin{document}
\title
{Quadrupole deformation in $\Lambda$-hypernuclei}
\author{Bipasha Bhowmick 
\thanks{bips.gini@gmail.com} \and
 Abhijit Bhattacharyya 
\thanks{abphy@caluniv.ac.in} \and
 G. Gangopadhyay
\thanks{ggphy@caluniv.ac.in}}
%\email
\institute{Department of Physics, University of Calcutta\\
92, Acharya Prafulla Chandra Road, Kolkata-700 009, India}

\abstract{
Shapes of light normal nuclei and $\Lambda$-hypernuclei are investigated using 
relativistic
mean field approach. The FSUGold parametrization is used for this purpose.
The addition of a $\Lambda$ is found to change the shape of the energy surface 
towards positive deformation {\em i.e.} prolate shape. The deformation in a 
$\Lambda$-hypernucleus, when the hyperon is in the first excited state, is also discussed.
The effect of the inclusion of the hyperon on the nuclear radius  
is generally small with one exception. 
}
\maketitle

\section{Introduction}
One of the unique and interesting aspects of hypernuclei is the structural 
change caused
by the hyperon. As an impurity in normal nuclei, a hyperon may be expected  to
induce
many effects on the core nucleus, such as change
in size\cite{Hiyama,Tanida}, shape change, modification of its cluster structure\cite{Hiyama2}, 
occurrence of nucleon and hyperon skin
or halo\cite{Hiyama2,Lu}, shift of neutron drip line to more
neutron-rich side\cite{Lu,Zhou,Bhowmick} etc.
Motoba \etal \cite{Motoba1,Motoba2} analyzed the light p-shell hypernuclei in the microscopic cluster
model and studied the binding energy of ground state, excited states and 
transition probabilities.
Owing to recent
experimental developments some of those have been observed in light p-shell hypernuclei.
As examples, we can refer to the reduction of B(E2) in $^{7}_{\Lambda}$Li\cite{Tanida} and the identification of the
super-symmetric hypernuclear state in $^{9}_{\Lambda}$Be\cite{Hashimoto}.
We can expect that a new experimental facility of Japan Proton Accelerator Research Complex (J-PARC)
will reveal new spectral information on p and sd shell as well as neutron-rich
hypernuclei.

As the shape of nuclei plays a decisive role in determining
 their properties, such as quadrupole moments and radii,
 mean field calculations have been performed
in recent years to investigate the change of nuclear shape
due to the addition of a $\Lambda$ hyperon. Deformed 
Skyrme-Hartree-Fock(SkHF) studies in Ref. \cite{Zhou2} have shown that
the deformation of any  hypernucleus is slightly less than 
the corresponding core nucleus.
On the other hand, relativistic
 mean field (RMF) study in Ref. \cite{Hagino} found that the 
deformation completely disappears in
$^{13}_{\Lambda}$C and $^{29}_{\Lambda}$Si hypernuclei though the corresponding
normal core
nuclei are deformed. 
 Recently, a study with anti-symmetrized molecular dynamics by Isaka {\it et al.}\cite{Isaka} have 
found that the $\Lambda$ in p-wave enhances the nuclear deformation, while that in s-wave reduces it.

To perform a systematic and quantitative study of the structure change of the 
p and sd-shell nuclei, caused by the hyperon, we
study the binding energy, quadrupole deformation, and the root mean square radii of a number 
of hypernuclei within a RMF model.
 The relativistic mean field theory 
of nucleus has been fairly successful in reproducing the properties of finite nuclear systems\cite{Ring}.
It has also been extensively applied to the study of hypernuclei\cite{Shen}.

\section{Model Calculation}

There have been a number of RMF parametrizations for prediction of the nuclear ground
state properties. 
 In the present work the FSUGold Lagrangian density
 has been employed\cite{todd}.
This parametrization has already been 
extended to include hyperons and to study
the properties of hypernuclear systems\cite{Bhowmick2}.
In this work it is used to study the change in quadrupole deformation
due to addition of hyperon. 

The parameters of the $\Lambda$N interaction 
have been determined by fitting the experimental separation energies of $12$ hypernuclei in the mass 
region $16$ to $208$\cite{Bhowmick2}. 
The masses of the two physical mesons
have been taken from experiment. The nucleon mass
is taken as 939 MeV. The 
mass of the $\Lambda$ has been fixed at 1115.6 MeV.

In the present work, we have chosen to limit the type of deformation to azimuthally symmetric and reflection symmetric 
systems which corresponds to prolate and oblate ellipsoids for quadrupole 
deformation. This eliminates all the 
three-vector components of the boson fields. 
Therefore, the limitation on the type of deformation simplifies the calculation with contributions from only
the scalar meson, the zero components of
the isoscalar vector meson, the photon, and the 
neutral $\rho$ meson fields. This is the same set
of boson fields that was required for spherical nuclei; however,
these boson fields now have an additional angular dependence.

Solving the field equations with no further simplification can be 
very involved due to the 
difficulties encountered in obtaining solutions for coupled partial differential
equations. 
We expand the boson fields in terms of Legendre polynomials and 
the nucleonic orbital wave functions  in terms of spherical angle functions. The method of solution 
has been explained in detail in Ref. \cite{Gangopadhyay}.
Terms up to angular momentum $L=6$ have been taken into account.
The grid size for solving the differential equations is taken to be 0.1 fm. 

Using density functional theory, H\={a}kkinen \etal\cite{PRL1} showed that 
light nuclei and clusters of alkali-metal atoms have similar shapes and 
odd-even staggering of total energy, which they found to be nearly independent 
of the interactions between the fermions. In very light nuclei, the spacings 
between energy levels are large. Hence, pairing is not expected to play a very 
important role. In a deformed RMF calculation, Arumugam \etal\cite{Aru} found that 
pairing is unimportant in light nuclei. Hence, in the present 
calculation, we have neglected pairing. In the case of $^{25}_\Lambda$Mg, we 
have checked that the deformation becomes slightly smaller in both the core and the hypernucleus on inclusion of the pairing. However, the relative change in 
deformation on inclusion of the hyperon is unaffected by pairing. 

As will be seen, it has been necessary to study the energy surface 
of the ground state as a function of quadrupole deformation.
To calculate the energy surface as a function of quadrupole
deformation, we have used the method of Lagrange's undetermined multiplier. 
Thus the deformation constrained Lagrangian density has been written as
\be \mathcal{L}'=\mathcal{L}-\lambda\hat{Q}_D\bar{\Psi}{\Psi},\ee
where $\hat{Q}_D=f(r) (3z^2-r^2)$ is the quadrupole operator multiplied
by a radial damping function $f(r)$. Lagrange's equations
have been obtained from eqn. (1) and numerical solutions have been obtained 
for them. Following the method outlined in \cite{Proj}, the value of 
$\lambda$ has been adjusted to obtain the given expectation value of 
$\hat{Q}_D$. Bassichis \etal\cite{Proj}  have shown that the lowest
energy solution is obtained for each $<\hat{Q}_D>$.
For each $Q_D$, we have calculated the quadrupole deformation parameter
$\beta$ from the charge quadrupole moment using the relation
\be Q_p=\sqrt{\frac{16\pi}{5}}\frac{3}{4\pi}ZR_0^2\beta\ee
where $R_0=1.2 A^{1/3}$ fm.

The proton root mean square (rms) radius is defined as
\be r_p=\sqrt{\frac{1}{Z}\int r^2\rho_p d{\mathbf r}}\ee
Here $\rho_p$ denotes proton density. Neutron rms radius ($r_n$) and nuclear rms
radius ($r$) are also calculated in an analogous way. 
The nuclear radius, in case of hypernuclei, includes the effect of the 
hyperon also in the 
sense that the density in the above equation includes proton, neutron and 
hyperon densities. 

\section{Results}

We calculate the 
quadrupole deformation parameter and the rms radii for a number of
$\Lambda$-hypernuclei up to $A<30$ putting the
$\Lambda$ in its deformed ground state and also in the first excited state,
which have opposite parities. 
The results of our calculation are presented in Table 1.

\begin{table}[ht]
\begin{center}
 \caption{Calculated Binding energy/nucleon(-E/A) and $\Lambda$ separation energy ($E_{\Lambda}$) in MeV, quadrupole deformation
 parameter $\beta$, and the rms radius in fm at the minimum energy
for the core nucleus and the corresponding hypernucleus 
in different $\Lambda$ states.} 
\label{tab:2}
\begin{tabular}{ccccccc}
\noalign{\smallskip}\hline
Nucleus &   -E/A   & $E_{\Lambda}$ &   $\beta$ & $r_{p}$ & $r_{n}$& $r$\\
 ($\Lambda$-state)\\ \hline

$^{8}$Be               & 5.668    &   -              &0.38   &2.41    &2.31    & 2.36  \\
&5.533 &&-0.37& 2.39& 2.28 &2.33\\
$^{9}_{\Lambda}$Be (s)    & 5.837    &   7. 189$^*$     &0.34      &2.40    &2.38    & 2.34  \\\hline
$^{10}$B         & 6.201     &  -             &0.27      &2.41    &2.26    &2.33   \\
&6.116&&-0.20&2.39&2.24&2.32\\
$^{11}_{\Lambda}$B (s)    & 6.501    &   9.501         &0.33     &2.54    &2.34    &2.40    \\\hline
$^{12}$C           & 7.175     &  -        &     -0.21     &2.36    &2.18    &2.27   \\
$^{13}_{\Lambda}$C (s)    & 7.534     &  11.842$^\dagger$   & 0.13      &2.59    &2.20    &2.39   \\
$^{13}_{\Lambda}$C (p)    & 6.683     &  0.779         &0.14      &2.61    &3.00    &2.92   \\\hline
$^{18}$F         & 7.562    &  -              &0.14     &2.70    &2.45    &2.58   \\
&7.561 & &-0.11 & 2.69 & 2.44 & 2.57\\
$^{19}_{\Lambda}$F (s)    & 7.883    &   13.661        &0.13      &2.71    &2.52    &2.56   \\
$^{19}_{\Lambda}$F (p)    & 7.399    &   4.465         &0.14      &2.71    &2.62    &2.63    \\\hline
$^{22}$Na           & 7.565    &  -         &     0.22      &2.85    &2.57    &2.72   \\
&7.475&&-0.13&2.83 & 2.55 & 2.69\\
$^{23}_{\Lambda}$Na (s)   & 7.904    &    15.362       &0.21      &2.86    &2.62    &2.70   \\
$^{23}_{\Lambda}$Na (p)    & 7.537    &    6.921        &0.22      &2.86    &2.69    &2.74    \\\hline
$^{24}$Mg         & 7.814    &  -         &     0.25      &2.91    &2.60    &2.76   \\
&7.640 & &-0.15 & 2.89 & 2.59&  2.74\\
$^{25}_{\Lambda}$Mg (s)   & 8.142    &    16.014       &0.23      &2.91    &2.65    &2.74   \\
$^{25}_{\Lambda}$Mg (p)    & 7.823    &     8.039       &0.24      &2.91    &2.71    &2.78    \\\hline
$^{25}$Mg           & 7.872    &  -         &     0.20      &2.88    &2.64    &2.76   \\
&7.793 & &-0.13 & 2.87 & 2.63 & 2.75\\
$^{26}_{\Lambda}$Mg (s)   & 8.248    &     17.648      &0.22      &2.92    &2.63    &2.75   \\
$^{26}_{\Lambda}$Mg (p)    & 7.927    &     9.30        &0.23      &2.92    &2.69    &2.79    \\\hline
$^{25}$Al            & 7.657    &  -         &     0.15      &2.95    &2.56    &2.77   \\
  &7.565&&-0.11& 2.94 & 2.55 & 2.76\\
$^{26}_{\Lambda}$Al (s)   & 8.001    &     16.601      &0.14      &2.95    &2.61    &2.75   \\
$^{26}_{\Lambda}$Al (p)    & 7.676    &     12.103      &0.15      &2.95    &2.67    &2.79    \\\hline
$^{26}$Al            & 7.828    &  -         &     0.11      &2.92    &2.60    &2.76   \\
&7.795 & &-0.09 & 2.92 & 2.59 & 2.76\\
$^{27}_{\Lambda}$Al (s)   & 8.200    &     17.872      &0.13      &2.96    &2.59    &2.76   \\
$^{27}_{\Lambda}$Al (p)    & 7.874    &      9.07       &0.14      &2.96    &2.65    &2.80    \\\hline
$^{28}$Si         & 8.080    &  -         &     0.00      &2.95    &2.54    &2.77   \\
$^{29}_{\Lambda}$Si (s)   & 8.453    &      18.897&     0.06      &3.00    &2.54    &2.78   \\
$^{29}_{\Lambda}$Si (p)    & 8.118    &      9.182      &0.06      &3.00    &2.60    &2.82    \\\hline
\end{tabular}
\end{center}
$^*$Exp. value is $6.71\pm0.04$ MeV\cite{Davis}

$^\dagger$Exp. value is 11.69$\pm0.12$ MeV\cite{Davis}%, 11.38 MeV\cite{Hashimoto}
\end{table}

In our calculation, the ground state 
of $\Lambda$ is a pure $s$-state as the higher positive parity states are
too high in the continuum. The first excited state is a negative parity state
which has contributions from both the $p$ orbitals. 
We thus indicate the solutions for hypernuclei with the hyperon in the ground 
or the first excited states with `s' and `p' in parentheses in the table and in the  relevant figures.
We have looked for both prolate as well as oblate minima in normal nuclei
and hypernuclei.
It is observed that in all the cases, the deformed hypernuclear minimum
corresponds only to prolate 
deformation. The lowest state in the  $p$-shell, the $p_{1/2}$ state, 
splits up in $K^\pi=1/2^-$ and $K^\pi=3/2^-$ Nilsson states  
in presence of reflection symmetric axial deformation. For positive
deformation, the former is the lowest energy state.
Thus, in case of the excited state, the hyperon occupies the 
$K^\pi=1/2^-$ state.
No  oblate minima are observed in the hypernuclear systems. This 
observation will be elaborated later in this work. Very little information 
is available 
for the $\Lambda$ separation energy in the ground state of the hypernucleus; the existing experimental values agree reasonably well with our calculation.

From Table 1, we see that when the $\Lambda$ is placed in the ground state,
the deformation changes slightly on inclusion of  the hyperon, with only one 
exception in the case of  $^{13}_{\Lambda}$C. The ground 
state of the core nucleus is generally prolate, with the exception of 
$^{12}$C and $^{28}$Si, the former of the two being oblate and the latter 
spherical.
Except in these two nuclei, the deformation, on inclusion of a hyperon in the lowest energy state, may 
increase as well as decrease, but in general 
the shape remains similar to that of the core nucleus in its ground state. 

However, in the case of $^{13}_{\Lambda}$C, the shape appears to change drastically 
on the inclusion of the $\Lambda$. The ground state of $^{12}$C shows negative 
deformation {\em i.e.} oblate shape, whereas the corresponding hypernucleus
$^{13}_{\Lambda}$C is prolate. This is consistent with the fact that the energy 
surface  plotted against quadrupole deformation (Fig. 1) shows an
oblate minimum in the core nuclei but not in the hypernuclei.
\begin{figure}[h]
\resizebox{8cm}{!}{
\includegraphics{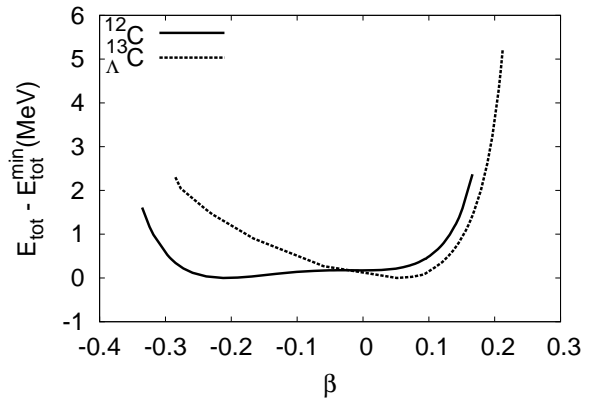}
}
\caption{Total energy with respect to the minimum (E$_{tot}$-E$^{min}_{tot}$) as 
function of deformation $\beta$ for $^{12}$C and the corresponding 
hypernucleus $^{13}_{\Lambda}$C. \label{c}
\label{fig:7}}
\end{figure}

This necessitated further investigation where we study the total  energy of the system against deformation in a constrained calculation.
In Fig. \ref{c} we
have represented the total binding energy with respect to the minimum 
(E$_{tot}$-E$_{tot}^{min})$ for $^{12}$C and $^{13}_{\Lambda}$C as a function of deformation
parameter $\beta$. We see that the energy surface of  $^{12}$C is almost flat 
with a oblate minimum around $\beta = -0.2$. Though there is no minimum for 
positive  deformation, around $\beta=0.1$ there exists a region which might
have formed a minimum.
Inclusion of the $\Lambda$
makes the oblate minimum disappear and the prolate minimum is found to be formed
in the energy surface of $^{13}_{\Lambda}$C.
One can see that although the unconstrained calculations predict 
the ground state of $^{12}$C and $^{13}_\Lambda$C to be oblate and prolate, 
respectively,  actually  a flat minimum is
being replaced by a sharper prolate one.
Our result is different from the previous HF calculations by Zhou~\etal\cite{Zhou2}
or the SkHF+BCS calculations by Win~\etal\cite{Win}, which have reported similar deformations for $^{12}$C and $^{13}_{\Lambda}$C.
Some calculations have also reported a spherical configuration for $^{13}_{\Lambda}$C\cite{Hagino,Isaka,Lu2}.

\begin{figure}[h]
\resizebox{8cm}{!}{
\includegraphics{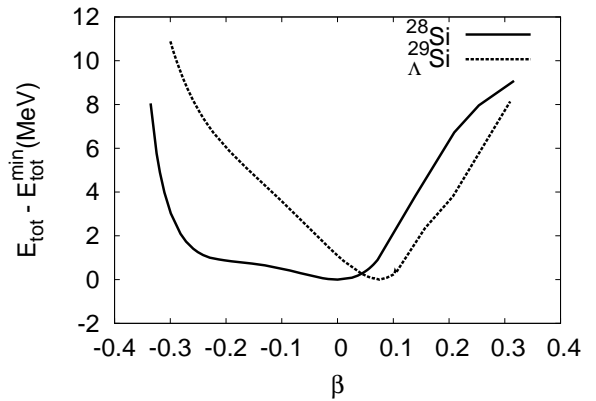}
}
\caption{Total energy with respect to the minimum as function of deformation 
$\beta$ for $^{28}$Si and the corresponding 
hypernucleus $^{29}_{\Lambda}$Si. \label{si}
\label{fig:8}}
\end{figure}

As a case where deformation appears in hypernuclei while the core is spherical 
in our calculation, we study $^{29}_\Lambda$Si.  However, we should point out 
that $^{28}$Si is experimentally known to be a deformed nucleus. 
%Lu~\etal\cite{Lu2} concluded that the prediction of the shape of $^{29}_{\Lambda}$Si is parameter dependent.
We see that $^{29}_{\Lambda}$Si is slightly prolate with the FSUGold parameter whereas $^{28}$Si is spherical.
In Fig. \ref{si} we
have represented the total binding energy with respect to the minimum 
$(E_{tot}-E_{tot}^{min})$ for $^{28}$Si and $^{29}_{\Lambda}$Si as a function of deformation.
From Fig.  \ref{si}, we see that the rather flat energy minima at the spherical
configuration in case of $^{28}$Si, becomes sharper and shifts
slightly towards positive side in case of $^{29}_{\Lambda}$Si.
One can also see that in the case of $^{28}$Si, a very slight change in the  
energy surface may make the ground state oblate consistent with the large deformation observed experimentally.

The situations in both the above cases may be considered as follows. The 
energy surfaces around the minima are rather flat, irrespective of the actual 
positions of the minima. 
The $1\Lambda s_{1/2}$ state, on the other hand, has a shallow prolate minimum.
The reason for the minimum occurring only on the prolate side requires further 
analysis which we plan to carry out in a future work. 
The inclusion of the $\Lambda-$hyperon in the $1s_{1/2}$ state is thus strong 
enough to polarize the core. Although individually the prolate minimum for the  
 $1\Lambda s_{1/2}$ state is not very deep, the interaction between the hyperon and the 
nucleons leads to a
change in the energy levels of normal nucleons also, driving the nucleus towards 
positive deformation.
Thus the overall surface shows a sharper prolate 
minimum. In case where the 
minimum is deep enough in the core nucleus, the hyperon does not change the 
shape.
For example, the nucleus $^{16}$O, being doubly magic, is known to be spherical
and has a deep minimum around  $\beta=0$.  We find that the corresponding 
hypernucleus $^{17}_\Lambda$O is also spherical.

We also note that our results for $^{25}_{\Lambda}$Mg agree with the 
SkHF+BCS calculations by Win~\etal\cite{Win} in the sense that inclusion of a
hyperon decreases the total deformation slightly. 
However, our results, as a  whole,  are different from the 
previous RMF and SkHF calculations\cite{Lu,Zhou,Zhou2,Hagino,Win,Lu2} in 
the sense that all the previous calculations report a decrease
in deformation on addition  of a $\Lambda$ particle. Our calculations 
show that the deformation may increase in some cases.  
In general, the inclusion of the $\Lambda$ hyperon makes
the energy surface sharper, the oblate minima disappears and the prolate one 
becomes prominent, thus producing a shift in deformation. 
Similar observations have been found for the other nuclei investigated in Table 1.

In the case of excited $\Lambda$ states, our calculations show that when the 
$\Lambda$ goes from the ground-state
to the p-state it slightly increases the deformation excluding the case of $^{29}_{\Lambda}$Si, where the deformation remains the same.
Calculation for the deformation of hypernuclei for an excited $\Lambda$ state 
reported by Isaka \etal\cite{Isaka} found similar results. In $^9_\Lambda$Be and $^{11}_\Lambda$B, the excited $\Lambda$ state is unbound.

We have also investigated the changes in the rms radii of the hypernuclei.  
Proton, neutron and total rms radii are
shown in the last three columns of Table 1. 

It is 
seen that the inclusion of a $\Lambda$
hyperon, causes  either a slight increase or no modification in neutron and 
proton radii when the $\Lambda$ particle is in 
the s-state. 
When the hyperon is excited to the p-state the size of the hypernucleus 
increases only slightly in almost all cases. The only exception is
$^{13}_{\Lambda}$C, where this increase in radius is significant, so that the hypernucleus is much larger
 than the corresponding core nucleus. The reason for this large increase 
is the 
change in shape. As already pointed out, the core in $^{13}_{\Lambda}$C is an oblate spheroid while 
the hypernucleus is prolate. The hypernucleus being elongated along the symmetry axis shows a larger radius.

We should point out that in contrast to many calculations, we have not
obtained any shrinkage in the ground state of the hypernucleus. 
We have already 
discussed the case of $^{13}_{\Lambda}$C. In $^{11}_\Lambda$B and 
$^{29}_\Lambda$Si, where the positive deformation has increased, we find a more 
significant increase in proton radius. Otherwise the proton radius is nearly 
equal to that of the core nucleus. In almost no case we have got a 
shrinkage. The neutron radius increases in almost all cases. 
A recent RMF calculation have also found no change in proton radius\cite{rmf}
though it should be added that the hypernuclei studied in that work are beyond 
the mass range of the present work.
The experimental evidence of shrinkage in charge radius is the decrease in 
B(E2) in $^7_\Lambda$Li, a very light nuclei observed by Tanida \etal
\cite{Tanida}. In our case, we get a slight decrease in the lightest nuclei,
$^9_\Lambda$Be.

 \begin{figure}[h]
\resizebox{8cm}{!}{
\includegraphics{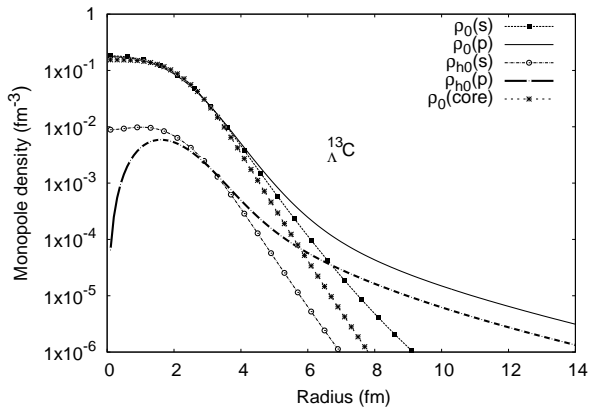}
}
\caption{The monopole nucleon and hyperon density profile in $^{13}_{\Lambda}$C when the $\Lambda$ is in different single particle 
state (s and p, given in parentheses). Here $\rho_o$ refer to monopole nucleon density and $\rho_{h0}$ refer to monopole hyperon density. The density of the
core nucleus $^{12}$C ($\rho_0$(core)) is also shown.  \label{1}.
\label{fig:1}}
\end{figure}

\begin{figure}[h]
\center
\resizebox{8cm}{!}{
\includegraphics{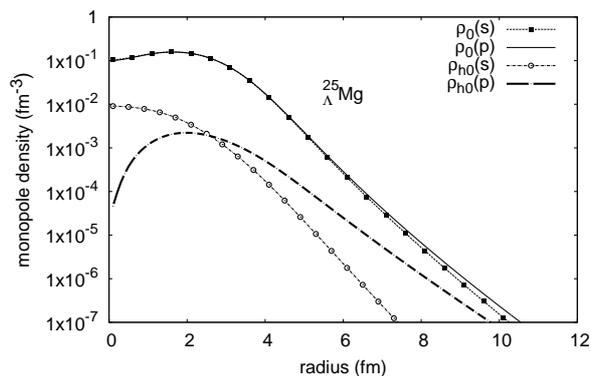}
}
\caption{
The monopole nucleon and hyperon density profile in $^{25}_{\Lambda}$Mg when the $\Lambda$ is in different single particle 
state (s and p). See caption of Fig. \ref{1} for details.
\label{3}}
\end{figure}

The reason for the large increase in size in $^{13}_\Lambda$C can be seen in the distribution of densities of the nuclei. We plot the monopole densities in
 $^{13}_{\Lambda}$C and $^{25}_{\Lambda}$Mg 
in Figs. \ref{1} and \ref{3}, respectively,
to see the effect of the hyperon. 
Both
the cases with the hyperon in its ground state and excited state are shown.  
In Fig. \ref{1}, the monopole density  of $^{12}$C is also shown. One can see 
that the density profile of this core nucleus differs from that of the normal 
nucleons in $^{13}_{\Lambda}$C, an evidence of the change in the ground state
as discussed earlier.

Fig. \ref{1} shows that 
in $^{13}_{\Lambda}$C,
when the hyperon is placed in the p-state, the densities extend to larger 
distances. Actually, the
hyperon is very loosely bound in the excited state. Hence, its wave function extends to a very large 
distance. As the hyperon interacts strongly with the nucleons, the nucleon density also shows 
a large tail, reminiscent of halo nuclei.

However, this is not a general phenomenon as evident from Table I. As an example, we 
present the density profile for $^{25}_\Lambda$Mg in Fig. \ref{3}. It is clear that in
this hypernucleus,  both the nucleon and the hyperon densities remain 
practically unchanged whether the hyperon is in the s-state or in the p-state. 
The p-shell hyperon in this case is strongly bound and does not show any 
halo-like behaviour.

\section{Summary}

We have studied the deformation of core and hypernuclei in the RMF approach using 
the FSUGold parameter set. The calculated $\Lambda$ binding energies agree 
reasonably well with the experimentally observed values. We see that the 
inclusion of a $\Lambda$ hyperon changes the energy surface making it steeper.  
There exists no oblate minimum in any light hypernucleus.  Results of the  present 
calculation differ from the previous ones. The latter usually predict that inclusion 
of a $\Lambda$ 
tends to drive the shape to spherical. Our results show that the change in $\beta$ is 
usually small with possible exceptions. If the core nucleus is prolate, the 
deformation may increase in some cases on addition of a hyperon.
In general, the nucleon density profile 
changes to a small extent  on inclusion of the hyperon, whether in the ground 
state or the first excited state. 
When the $\Lambda$ goes to the p-state, both the deformation and radius 
increases by
a small amount. The only exception is $^{13}_\Lambda$C where, the hyperon being very loosely
bound, creates a halo-like structure.

\begin{acknowledgement}

This work was carried out with financial assistance of the UGC (UPE, RFSMS, DRS) and DST of the Government of India. 

\end{acknowledgement}

\end{document}